\documentstyle[12pt]{article}
\renewcommand {\d}  {\hbox{d\hskip-1.1ex{\raise0.640ex\hbox{--}}\skip 0.70ex}}
\newcommand   {\D}  {\hbox{D\hskip-1.9ex{\raise0.175ex\hbox{--}}\hskip0.85ex}}

\begin{document}

\newcounter{buco}
\setcounter{equation}{0}
\setcounter{buco}{0}
\renewcommand{\thesection}{\Roman{section}}
\renewcommand{\theequation}{\arabic{section}\arabic{equation}}

\def\theequation{\thesection.\arabic{equation}}
\def\theeqnarray{\thesection.\arabic{eqnarray}}

\begin{center}
\begin{Large}
\begin{bf}
 Covariant and heavy quark symmetric quark models \\
\end{bf}
\end{Large}
\vspace{2.0cm}
D. Tadi\' c and S. \v Zganec \\
\vspace{1.5cm}
Physics Department, University of Zagreb, Bijeni\v cka c. 32, Zagreb,
Croatia
\end{center}
\vskip 2cm
\begin{center}
\bf Abstract
\end{center}
\baselineskip=24pt

\noindent
There exist relativistic quark models (potential or MIT-bag)
which satisfy the heavy quark symmetry (HQS) relations among meson decay
constants and form factors. Covariant construction of the momentum
eigenstates, developed here, can correct for spurious center-of-mass motion
contributions.
Proton form factor and M1 transitions in quarkonia are calculated. Explicit
expression for the Isgur-Wise function is found and model determined
deviations from HQS are studied. All results depend on the model parameters
only. No additional ad hoc assumptions are needed.

\newpage

\section{Introduction}
\vspace{2cm}

\indent

A simple, but covariant quark model[1-4], used previously to
calculate meson form factors[5], posesses  also the heavy quark symmetry
$(HQS)$[6-14]. Actually this might be true for a whole class of quark models.
This class contains models in which quarks are confined by a central
potential. Their wave functions must be Lorentz boosted [2-5] It is hoped
that such models might serve as a useful semiempirical tool. They can be
used to roughly estimate physical quantities and effects and to illustrate
HQS relations.

Once the model
confinement parameters [15-17], the  quark masses and the
interaction hypersurface[3,5]  are selected, everything else follows from
our formalism. No additional assumptions, as for example about
$Q^{2}$-dependence of form factors [18]  are needed. The HQS is intimately
connected with the Lorentz-covariant character of the model.

Models hadron states, used previously [1-3], were not momentum
eigenstates [19-24]. This can be remedied by a
projection  [19-25] of model states into momentum eigenstates. A Lorentz -
covariant
projection [25] is developed here. It is shown
that this removal of the spurious center-of-mass  motion improoves the
model description of proton electromagnetic form factors. Such corrections
are not important if hadron contains heavy quarks c or b. In that case they
are smaller than 5 \%.

The model calculations give some corrections to the extreme HQS.
Some of those, for example concerning meson decay constants $f_{D}$ and
$f_{D_{s}}$ agree with QCD sum  rule results [26]. Model predictions for
meson form factors in the heavy
quark limit (HQL) follow exactly the HQS requirements. One can
extract model prediction for the Isgur-Wise function  $\xi$ [7].

\newpage

\section{Relativistic model}
\vspace{2cm}
\setcounter{equation}{0}
\renewcommand{\theequation}{\arabic{section}.\arabic{equation}}

\indent

Any static model in which quarks are confined  by a central force
can be relativized [1-5]. Earlier the MIT-bag model has been employed [3-5].
Here a harmonic oscillator confining potential [16-17] will be used.

 In any of them one can
envisage a hadron as located around $y$. The quark $q_{i}$ coordinate is
\begin{equation}
x_{i}=y+z_{i}
\end{equation}

The confining "ball" of mass M, can be boosted,
acquireing the four-momentum P.
Individual quark wave functions $\psi_{n}$ depend on z and P
\begin{equation}
\psi_{n}(z^{P})=S(P)\eta_{n}(Z^{P}_{\perp})exp(-iz^{P}_{\parallel}\epsilon_{n})
\end{equation}

Here
\begin{eqnarray}
S(P) & = & (\not \! P \gamma_{0}+M)/[2M(E+M)]^{1/2} \nonumber \\
E & = & (\vec {P}^{2}+M^{2})^{1/2}  \nonumber \\
z_{\perp}(P)_{\mu}  & = & z_{\mu}-\beta^{\mu}(\beta \cdot z)  \nonumber \\
z_{\parallel}(P) & = & \beta_{\mu}z^{\mu}  \nonumber \\
\beta_{\mu} & = & P_{\mu}/M
\end{eqnarray}
and $\epsilon_{n}$ is model energy.
For $\beta_{\mu}=0$ the Dirac spinor $\eta$ has a generic form
\begin{equation}
\eta_{n}(\vec {r})= \left (
\begin{array}{c}
U_{n}(|\vec{r}|)\chi\\
i\vec{\sigma}\frac {\vec{r}}{|\vec{r}|} V_{n}(|\vec{r}|)\chi
\end{array}
\right )
\end{equation}
Here $\chi$ is the Pauli spinor.

One can introduce the quark field operator
\begin{equation}
\Psi(z^{P})=\sum_{n}(a_{n}\psi_{n}(z_{P})+b^{*}_{n}\overline
{\psi_{n}}(z_{P}))
\end{equation}
and define model states for meson "m" for example:
\begin{equation}
| m,M,P,s,y \rangle=\sum_{r,r',f,f'} C^{s}_{r r^{'}f f^{'}}
a^{*rf}_{nP} b^{* r^{'} f^{'}}_{n P} | 0 \rangle e^{-iPy}
\end{equation}
Here $m$ is the flavor (B,D etc.),$M$ is the meson mass and $s$  is the spin.

Using the configuration space operators [27] (2.5) one can obtain
a model wave functions whose generic form is:

\begin{eqnarray}
\langle 0|\Psi(z_{1}^{P})\Psi(z^{P}_{2})\Psi(z^{P}_{3})|b,M,P,s,y \rangle
& = & N_{P}e^{-iPy}F_{b}(\psi_{f_{1}}(z^{P}_{1})\psi_{f_{2}}(z^{P}_{2})
\psi_{f_{3}}(z^{P}_{3})) \nonumber \\
& = & h^{s}_{b}(P, z_{1}, z_{2}, z_{3}) e^{-iPy}
\end{eqnarray}
        Here $N_{P}$ is the norm and $F_{b}$ symbolizes the symmetrized
combination of quark flavors.

A quark line in the configuration
space, in the non relativistic limit, corresponds to the normalization
integral
\begin{equation}
\int \psi^{*} \psi d^{3} z = 1
\end{equation}
This can be generalized as
\begin{equation}
Z=J(z)\bar{\psi}(z^{P_{f}})\not \! L \psi(z^{P_{i}})
\end{equation}
Here
\begin{equation}
J(z)=\int d^{4}z \delta (L z)
\end{equation}
Among all possible hypersurfaces
\begin{equation}
L\cdot z = 0
\end{equation}
only the one defined by
\begin{equation}
L_{\mu}=(\beta^{i}_{\mu}+\beta^{f}_{\mu})/[(\beta_{i}+\beta_{f})^{2}]^{1/2}
\end{equation}
leads to the proton electromagnetic formfactors $f_{i}$ which satisfy the
conserved current constraint $f_{3}(Q^{2})\equiv 0$ (5.2). A model defined
on a hyperplane is connected [2,3]  with the quasipotential
approximation [28]  of the Bethe-Salpeter equation.

The vertex spatial dependence follows from $(2.9)$ by replacing
\begin{equation}
\begin{array}{c}
\not \! L \rightarrow \Gamma_{\mu}\\
\Gamma_{\mu} = \gamma_{\mu},\  \gamma_{\mu}\gamma_{5} ,\ etc.
\end{array}
\end{equation}
For mesons $m_{f},\ m_{i}\ (2.6)$ a current matrix element is
\begin{displaymath}
V(\Gamma^{\mu})=\int d^{4} y \langle y,s_{f},P_{f},M_{f},m_{f}|
J(z_{1}) \overline{\Psi} (z^{P_{2}}_{1}) \Gamma^{\mu} \Psi(z^{P_{1}}_{1})
\end{displaymath}
\begin{equation}
\cdot J(z_{2}) \overline{\Psi} (z^{P_{4}}_{2})\not \! L \Psi(z^{P_{3}}_{2})
|m_{i},M_{i},P_{i},s_{i},y \rangle
\end{equation}

\newpage

\section{Momentum eigenstates}
\vspace{2cm}
\setcounter{equation}{0}
\renewcommand{\theequation}{\arabic{section}.\arabic{equation}}

\indent

The factor $\exp(-iPy)$ (2.6) describes the motion of
center-of-force (CF).
The center-of-mass (CM) of
centrally confined quarks oscillates about CF. As it is well
known [1-6,25]  the spurious center-of-mass-motion (CMM) persist even in the
static
$(\vec{P}=0)$ case. Thus the boosted centrally confined model (BCCM)
 states (2.6) are not the momentum eigenstates.
This can be remedied by decomposing a BCCM state into momentum eigenstates
$|l,s \rangle$ as follows [19-24]:

\begin{eqnarray}
|h,M,P,s,y \rangle &=& 2M \int d^{4}l \delta (l^{2}-M^{2}) \theta (\omega)
 e^{-ily}\phi_{P}(l)|l,s\rangle = \nonumber \\
&=& 2M \int \frac{d^{3}l}{2\omega} e^{-ily} \phi_{P}(l)|l,s\rangle
\end{eqnarray}
Here $h$ denotes a hadron. The momentum eigenstates normalization is
\begin{eqnarray}
\langle l^{'}s^{'}|ls\rangle &=& \delta_{s s'}\delta (\vec {l}^{'}
 - \vec {l}) \nonumber \\
l&=&(\omega, \vec {l})
\end{eqnarray}
For a BCCM state one has
\begin{equation}
\langle h,M,P,s,0 | 0,s,P,M,h \rangle = 1
= \int d^{3}l \frac{M^{2}}{\omega^{2}} | \phi_{P}(l)|^{2}
\end{equation}
This provides a normalization of the components $\phi$  of the momentum
eigenstates.

The momentum eigenstates in (3.1) are not the exact physical hadron
states but the model hadron states, i.e. some kind of "mock" hadron
states [29].

In the occupation number space one finds for a baryon b, for example
\begin{equation}
\langle y = 0,s,P,M,b | b,M P,s,y = \zeta_{\perp} (P) \rangle =
M^{2}\int \frac {d^{3}l}{\omega^{2} } | \phi_{P} (\vec {l}, \omega)
|^{2} e^{-i l \zeta_{\perp}(P)}
\end{equation}
In the coordinate space this becomes
\begin{equation}
\begin{array}{c}
\langle y= 0,s,P,M,b|b,M,P,s,y = \zeta_{\perp}(P)\rangle =
 {\cal M}(P,\zeta_{\perp}(P)) = \nonumber \\
= J(z_{1})J(z_{2})J(z_{3}) h^{* s}_{b} (P,z_{1}, z_{2}, z_{3}) \not \! L_{1}
\not \! L_{2} \not \! L_{3}\cdot \\
\cdot h^{s}_{b}(P, z_{1}-\zeta_{\perp}(P), z_{2}-\zeta_{\perp}(P),
z_{3}-\zeta_{\perp}(P)) \\
\end{array}
\end{equation}
For the proton, with all light quark masses equal $(m_{u}=m_{d})$, one finds
\begin{equation}
{\cal M}(P,\zeta_{\perp}(P))=
[J(z)\overline{\psi} (z_{\perp}(P))S^{-1}(-\frac {\vec {P}}{E})\not \! L S
(-\frac {\vec {P}}{E}) \psi ((z(P)-\zeta (P))_{\perp})]^{3}
\end{equation}
Integrating (3.4) and (3.6) over $\zeta$  one finds
\begin{equation}
J(\zeta){\cal M}(P,\zeta_{\perp}(P))e^{il \zeta_{\perp}(P)}
= M^{2}J(\zeta)\int \frac {d^{3}k}{\omega^{2}_{k}}|\phi_{P}
(k,\omega_{k})|^{2} e^{i (l-k)\zeta_{\perp}(P)}
\end{equation}
The end result is the Lorentz-covariant expression for the components of the
momentum eigenstates:
\begin{equation}
\frac {M}{\omega_{l}} |\phi_{P}(\vec {l}, \omega_{l})|^{2}
=\frac {l\cdot P}{(2\pi)^{3} M^{2}}\int d^{4}\zeta\delta (L\cdot \zeta)
{\cal M} (P, \frac {\vec {P}\cdot\vec {\zeta}}{E}, \vec {\zeta})
e^{i l \zeta_{\perp} (P)}
\end{equation}
Some explicit expressions for $\phi's$ are listed in Appendix.

\newpage

\section{Confinement}
\vspace{2cm}
\setcounter{equation}{0}
\renewcommand{\theequation}{\arabic{section}.\arabic{equation}}

\indent

The Dirac equation for quarks can be solved for the potential
\begin {equation}
V(r,P) = \frac {1}{2} (1+\frac {\not \! P}{M}) (V_{0}-\frac {1}{2} K z_{\perp}
(P)^{2})
\end {equation}
which in the hadron rest frame has the harmonic oscillator (HO)
shape [16]
\begin {equation}
V(r) = \frac {1}{2} (1+\gamma^{0}) (V_{0}+\frac {1}{2}K r^{2})
\end {equation}
Here $V_{0}$ and $K$  are model parameters. The rest frame solution has a
general form (2.4), with:
\begin{equation}
\begin{array}{c}
U_{a} = exp (-r^{2}/2 R^{2}_{0 a}) \\
V_{a} = r \beta_{a}U_{a}/R_{0 a} \\
N_{a} = [R^{3}_{0 a} \pi^{3/2} (1+\frac {3}{2} \beta^{2}_{a})]^{-1/2} \\
\end{array}
\end{equation}
The index a denotes the quark's flavor. The quantities $R_{0 a}$ and
$\beta_{a}$  depend on the constituent mass $m_{a}$  and the energy $E_{a}$.
\begin{equation}
\begin{array}{c}
E_{a}=m_{a} + V_{0} + 3 [K/2 (m_{a}+E_{a})]^{1/2} \\ \nonumber
R^{4}_{0 a} = 2/K (m_{a} + E_{a}) \\
\beta_{a} = R^{-1}_{0 a} (m_{a}+E_{a})^{-1} \\
\end{array}
\end{equation}

An approximate solution [6]  for the linear potential
$V(r)=\frac {1}{2}(1+\gamma^{0})(V_{0}+\lambda r)$
would also have the form (4.3), with accuracy of $\sim 6$\%.  All general
HQS features, discussed  bellow would, thus apply for that potential also.

In the heavy quark limit (HQL),where $m_{a} \rightarrow \infty $ and
$E_{a}\rightarrow  m_{a}$, one has
\begin {equation}
\frac {\beta_{a}}{R_{0 a}} \rightarrow \frac {1}{2} \sqrt \frac
{K}{m_{a}}\rightarrow 0
\end {equation}
Thus only the "large" component $U$  survives in (4.3).

One can also show
that in MIT - bag model [15] "small" component $V$  vanishes in HQL.
In the numerical evaluation MIT - bag model parameters employed
previously by Ref.[5] will be used.

The HO model parameters are
\begin{equation}
\begin{array}{c}
V_{o} = -0.35 GeV \\ \nonumber
K= 0.035 GeV^{3} \\
\end{array}
\end{equation}
The constituent quark masses  and related quantities $\beta, E$  and $R_{0}$
are listed in Table~I.

Table II shows model hadron masses calculated using either model
states (2.6) or model dependent momentum eigenstates (3.2).
The relevant formula for the valence  quark contribution to the
hadron mass $\tilde{M}_{Q}$ is:
\begin{equation}
\begin{array}{c}
\tilde{M}_{Q}^{h} = \langle h M,0,s,0 | \int T^{00} d^{3}x|h,M,0,s,0\rangle
\\ \nonumber
= \langle h,M,0,s,0 | P^{0} | h,M,0,s,0 \rangle \\
\end{array}
\end{equation}
Here $T^{00}$ is the momentum energy tensor. One must add magnetic
$\Delta\tilde{M}_{M}$ and electric $\Delta \tilde{M}_{E}$  effective one gluon
exchange contributions [15,16]
which for the HO potential model can be calucated explicitly. Finally one
has BCCM based hadron mass without CMM corrections.
\begin {equation}
\tilde{M}=\tilde{M}_{Q}+\Delta \tilde{M}_{M}+\Delta \tilde{M}_{E}
\end {equation}
Using momentum eigenstates one obtains the following identities for a
meson $m$ or a baryon b:
\begin {equation}
\tilde{M}^{m} = \int \frac {d^{3}k}{4 \omega^{2}} | \varphi^{m}(k) |^{2}
\sqrt {M^{m^{2}}+\vec {k^{2}}}
\end {equation}
\begin{equation}
\tilde{M}^{b} = \int d^{3}k \frac {M^{b^{2}}}{\omega^{2}} | \phi^{b} (k)
|^{2} \sqrt {M^{b^{2}}+\vec {k^{2}}}
\end{equation}
Here $\tilde {M}'s$  and $\phi's$ are determined by parameters from
Table I. The CMM corrected masses $ M^{m,b}$  can be found numerically.
Inspection of Table II reveals that CMM corrections improove the agreement
with the experimental values[6]. The mass of the pion is quite wrong, as
in all valence quark models which do not account for the Goldstone - boson
nature of pion. Other theoretical masses are correct within 10\% or better.
CMM corrections increase mass difference in a  SU(6) multiplet, $(p,\Delta,
 etc.)$, bringing theory closer to experiment. Corrections decrease with the
increase of the heavy quark mass. Thus for example
$(\tilde{M}_{B}-M_{B})/M_{B}\cong 1.6 $\%.

\newpage

\section{Proton formfactors}
\vspace{2cm}
\setcounter{equation}{0}
\renewcommand{\theequation}{\arabic{section}.\arabic{equation}}

\indent

Calculation of the proton formfactors is a useful test of any
quark model.
All calculational detailes have been discussed and described
in Ref.'s [3] and [5]. It remains to be shown that the inclusion of CMM
corrections improoves upon earlier results.

These corrections are included by the equality
\begin{displaymath}
\int d^{4}y  \prod_{i=1}^{3} J(z_{i})  \langle M, P_{f},y |\sum_{i,j,k,perm}^{}
V^{\mu}(z_{i})C(z_{j})C(z_{k})\cdot e^{-iQ x_{i}} | M, P_{i}, y \rangle =
\end{displaymath}
\begin{equation}
 = (2\pi)^{4} \delta (P_{f}-P_{i}-Q) J(z)\int \frac {d^{3} l d^{3} l^{'}}
{\omega \omega^{'}} M^{2} \phi^{*}_{P_{f}} (\vec {l}^{'}, \omega^{'})
\phi_{P_{i}} (\vec {l}, \omega)
\langle \vec {l}^{'}|V^{\mu}(z) e^{-i Q z} | \vec {l} \rangle
\end{equation}

Here:
\begin{displaymath}
V^{\mu}(z_{i}) = \overline{\Psi}(z_{i}) \gamma^{\mu} \Psi (z_{i})
\end{displaymath}
\begin{equation}
C(z_{k})  =  \overline{\Psi} (z_{k}) \not \! L \Psi (z_{k})
\end{equation}
\begin{displaymath}
\langle l^{'}|V^{\mu}(0)|l\rangle  =  \bar{u}(l^{'}) [f_{1}(s^{2})
\gamma^{\mu} + f_{2}(s^{2}) i \sigma^{\mu \nu} s_{\nu} + f_{3}(s^{2})s^{\mu}]
u(l)
\end{displaymath}
\begin{displaymath}
s = l^{'}-l \ \ \ \ ; \ \ \  f_{3}(s^{2}) \equiv  0
\end{displaymath}
The l.h.s. of (5.1) is the expression used earlier [5] to calculate
electromagnetic formfactors. Here it is written in the occupation number
space.

In general one cannot invert the expression (5.1).
However at the momentum transfer
$Q^{2} = O$  one can determine [23] the Sach's form factor $G_{M}(0)$.

The l.h.s. of (5.1) can be written as

\begin{equation}
(5.1)(l.h.s.)=(2\pi)^{4}\delta(P_{f}-P_{i}-Q)\chi^{+}[W^{0}+\frac {i}
{2M}\vec {\sigma}\times \vec {Q}W^{2}]\chi
\end {equation}
Here
\begin{displaymath}
W^{0}=I_{0}\cdot Z^{2}
\end{displaymath}
\begin{displaymath}
W^{2}=I_{2}\cdot Z^{2}
\end{displaymath}
\begin{displaymath}
Z=\frac{M_{f}}{E_{f}} 4\pi \int dr r^{2} j_{0}
(\rho)[U^{2}+V^{2}]
\end{displaymath}

\begin{displaymath}
I_{0}=4\pi \frac{M_{f}}{E_{f}}\int dr r^{2} j_{0}(\tilde{\rho})
[U^{2}+V^{2}]
\end{displaymath}
\begin{displaymath}
I_{2}=4\pi \frac{M_{f}}{E_{f}}\int dr r^{2}[j_{0}(\tilde{\rho})U^{2}-(\frac
{1}{3}j_{0}(\tilde{\rho})-\frac{2}{3} j_{2} (\tilde{\rho}))V^{2} \\
 + \frac{2E_{f}}{|\vec{P_{f}}|}j_{1}(\tilde {\rho}) UV)]
\end{displaymath}
\begin{displaymath}
\rho = \frac{|\vec{P_{f}}|}{E_{f}} 2 \epsilon |\vec {r}|  \ \ \ \ ;\ \ \ \
\tilde {\rho}=\frac{|\vec{P_{f}}|}{E_{f}} 2 (M - \epsilon ) |\vec {r}|
\end{displaymath}
The quantities $W^{\alpha}$ were identified [3,5] as Sach's formfactors
\begin{equation}
\begin{array}{c}
W^{0}\sim G_{E} \\ \nonumber
W^{2}\sim G_{M} \\
\end{array}
\end{equation}
However (5.4) was obtained using BCCM states which are
not momentum eigenstates.

More accurate approach is based on the equality
\begin{displaymath}
(5.1)(r.h.s.)=(2\pi)^{4}\delta(P_{f}-P_{i}-Q)D^{\mu}
\end{displaymath}
\begin{displaymath}
D^{\mu}=2\pi \int l^{2}dl sin \theta d \theta \frac {M^{3}}{(\omega
\omega^{'})^{3/2}} \phi^{*}_{P_{f}}(\vec {l}, \omega^{'})\phi_{P_{i}}
(\vec {l}, \omega)
\frac {1}{\sqrt {4M^{2}(\omega+M)(\omega^{'}+M)}} \cdot
 \end{displaymath}
\begin{equation}
\frac {1}{(1-\frac {q^{2}}{4M^{2}})} [G_{E}(q^{2})
(\delta^{\mu}-\frac {\eta^{\mu}}{2M}) +
G_{M}(q^{2})(\frac {\eta^{\mu}}{2M} - \frac {q^{2}}
{4M^{2}}\delta^{\mu})] \cdot \chi^{+}\Gamma(\mu)\chi
\end{equation}
\begin{displaymath}
\chi^{+}\Gamma(\mu)\chi = \chi^{+}[\delta_{\mu 0} + \delta_{\mu 3} +
\delta_{\mu 1} i \sigma_{2}- \delta_{\mu 2} i \sigma_{1}] \chi
\end{displaymath}

Here
\begin{displaymath}
\vec {l^{'}}=\vec {l}+\vec {Q} \ \ \ \  ;\ \  \omega^{'2} = \vec {l^{'2}}
+M^{2} \\
\end{displaymath}
\begin{displaymath}
q = (q^{0}, \vec {Q}) \ \ \ \  ;\ \  q^{0} = \omega^{'}-\omega \\
\end{displaymath}
\begin{displaymath}
\delta^{0}=a_{i}a_{f} + \vec {l}^{2} + \vec {Q}\cdot \vec {l} \\
\end{displaymath}
\begin{displaymath}
\delta^{1}=\delta^{2}=(a_{i}-a_{f}) | \vec {l} | cos \theta + a_{i}|\vec{Q}|\\
\end{displaymath}
\begin{displaymath}
\delta^{3}=a_{f}|\vec {l}| cos \theta + a_{i}(|\vec {l}| cos \theta + |\vec
{Q}| )\\
\end{displaymath}
\begin{equation}
\eta^{0} = (a_{f}-a_{i}) \vec {Q} \cdot \vec {l} - a_{i} \vec {Q}^{2} \\
\end{equation}
\begin{displaymath}
\eta^{1}=\eta^{2}=a_{i}a_{f} |\vec {Q}| + |\vec {Q}| (\vec {l}^{2} +
\vec {l} \vec {Q}) - \vec {l}^{2} |\vec {Q}| sin^{2} \theta + \\ \nonumber
+ (\omega^{'}-\omega) (-a_{f}|\vec {l}| cos \theta - a_{i}|\vec {l}|
cos \theta - a_{i} |\vec {Q}|) \\
\end{displaymath}
\begin{displaymath}
\eta^{3} = (\omega^{'}-\omega) [a_{f}|\vec {l}| cos \theta - a_{i}(|\vec {l}|
 cos \theta + \vec {|Q|})] \\
\end{displaymath}
\begin{displaymath}
a_{i}=\omega + M \ \ \ \ ; \ \  a_{f} = \omega^{'} + M \\
\end{displaymath}
\begin{displaymath}
\vec {Q} \cdot \vec {l} = |\vec {Q}| |\vec {l}| cos \theta \\
\end{displaymath}

Four - momentum $q$  is an averrage value of $l'- l$  calculated between
two wave-packets $\phi_{P}$  which have speeds $\beta^{\mu}_{i}$ and
$\beta^{\mu}_{f}$ respectively.

For the Sach's form factors one can assume the well known dipole
shapes
\begin{equation}
\frac{G_{E}(q^{2})}{G_{E}(0)} = \frac {G_{M}(q^{2})}{G_{M}(0)}=
(1-\frac {q^{2}}{\eta^{2}})^{-1}
\end{equation}
The magnetic moment $G_{M}(0)=\mu_{P}$ can be determined from the equalities
(5.4) and (5.6) taken at $Q^{2} = O$. One obtains
\begin{displaymath}
\vec{D}|_{\vec{Q}=0}=0
\end{displaymath}
\begin{equation}
\vec{Q}\frac{\partial \vec {D}}{\partial \vec {Q}}|_{\vec {Q}=0} =\frac {1}{2M}
\chi^{+} i \vec {\sigma} \times \vec {Q} \chi \cdot \kappa =
\frac{1}{2M} \chi^{+} i \vec{\sigma} \times Q \chi \cdot W^{2}(\vec {Q}=0)
\end{equation}
\begin{displaymath}
\kappa=W^{2}(\vec {Q}=0)= 4\pi\int l^{2}d l \frac {M^{2}}{\omega^{2}}|
\phi_{\vec {P}=0} (\vec {l},\vec {\omega})|^{2}
[ \frac {G_{E}(0)}{3}(\frac {M}{\omega}-1)+ \frac {G_{M}(0)}{3} (1+
\frac {M}{\omega}+\frac {M^{2}}{\omega^{2}})]
\end{displaymath}
With  $G_{E}(0)=1$ one finds
\begin{equation}
G_{M}(0)=2.212
\end{equation}
which is about 20\% to small. However without CMM corrections one would
have obtained
\begin{equation}
G_{M}(0)=1.738
\end{equation}
which is much smaller than the experimental value [30] $G_{M}(0)=2.793$.
The CMM corrections have resulted in 27 \% improovement of the model
value[23].

The equality (5.5) can be used to determine the parameter $\eta$.
For $Q^{2}< 1.17GeV^{2}$  equality is, within 10\% error, satisfied with
\begin {equation}
\eta=0.70 GeV^{2}
\end {equation}
which is very close to the experimental value
[30] $\eta_{exp}=0.71 GeV^{2}$.

The fit (5.11) fails progresively as $Q^{2}$  increases above $1.17GeV^{2}$.
Qualitatively this agrees with other model based calculations, see for example
Ref. [31].

An analogous formalism can be used for the nucleon axial vector
coupling constant $g_{A}$. Without CMM corrections one finds $g_{A}=1.14$.
With CMM corrections the theoretical result $g_{A}=1.22$ is surprisingly
close to the experimental value [30].

A strong point in favor of BCCM with CMM corrections is that corrections
are much larger for $G_{M}$ (27\%) than for $g_{A}$ $(7\%)$, just as
needed.

\newpage

\section{M1 transition in quarkonia}
\vspace{2cm}
\setcounter{equation}{0}
\renewcommand{\theequation}{\arabic{section}.\arabic{equation}}

\indent

The M1 transitions

\begin{equation}
\begin{array}{c}
V\rightarrow P+\gamma \\ \nonumber
(^{3}S_{1}\rightarrow\ ^{1}S_{0} + \gamma) \\
\end{array}
\end{equation}
provide useful informations [32,33] about CMM corrections for systems
containing heavy quarks c and b.
The decay amplitude is
\begin{equation}
\langle m (P_{f})|J^{\mu}_{el.mg.}(0)|v (P_{i},\epsilon)\rangle
= \frac {1}{(2\pi)^{3}\sqrt {4E_{i}E_{f}}} g \epsilon^{\mu \nu \sigma \rho}
\epsilon_{\nu} (P_{f}-P_{i})_{\sigma}(P_{f}+P_{i})_{\rho} \\
\end{equation}
with the corresponding decay width
\begin{equation}
\begin{array}{c}
\Gamma(v\rightarrow m \gamma)=\frac {4}{3} \alpha (\frac {g}{l})^{2}
\omega_{\gamma}^{3} \\ \nonumber
\omega_{\gamma} = \frac {M^{2}_{i}-M^{2}_{f}} {2 M_{i}}
\end{array}
\end{equation}
Here $\alpha$ is the fine structure constant.

In BCCM's the form factor $g$ can be calculated with $(g(s^{2}))$ and
 without $(\tilde {g}(s^{2}))$ CMM corrections. In the first case one starts
with
\begin{eqnarray}
(2\pi)^{4}\delta(P_{f}+k-P_{i})&\cdot& \prod^{2}_{i=1} J(z_{i}) \sum_{l,n,perm}
\overline {\kappa_{m}}^{s=0}(P_{f},z^{P}_{1},z^{P}_{2},y=0) \nonumber \\
\cdot (\gamma^{\mu})_{l} e^{ikz_{l}}\not \! L_{n}
\kappa_{v}^{s=1}(P_{i},z^{P}_{1},
z^{P}_{2},y=0)&=&(2\pi)^{4}\delta (P_{f}+k-P_{i}) {\cal N}^{\mu} N_{fi} \\
N_{fi}&=&\frac{1}{(2\pi)^{3}} \sqrt{\frac{M_{i}M_{f}}{E_{i}E_{f}}} \nonumber
\end{eqnarray}

As ${\cal N}^{\mu}$ must have the same form as (6.2) one can identify the form
factor $\tilde {g}(0)$. Here $\kappa^{s}_{m}$ is the meson wave function
analogous to (2.7). The calculation was carried out in the generalized Breit
frame
\newpage
\begin{equation}
\frac {E_{i}}{M_{i}} = \frac {E_{f}}{M_{f}}\ \ \ ;\  \frac {\vec
{P_{i}}}{M_{i}} =
 - \frac {\vec {P_{f}}}{M_{f}} \ \ \ ;\  P_{i}=P_{f}+k
\end{equation}
\begin{displaymath}
|\vec {k}| = \frac {M^{2}_{i}-M^{2}_{f}}{2 \sqrt {M_{i}M_{f}}} \ \ \ ;\ \ \
|\vec {P_{i}}|=\sqrt {\frac {M_{i}}{M_{f}}} \frac {M_{i}-M_{f}}{2}
\end{displaymath}

By expansion of $\cal {N}^{\mu}$  around $Q^{2}=O$ one can find for
smaller $Q^{2}$
\begin{equation}
\tilde{g}(Q^{2}) = \frac{\tilde{g}(0)}{1-Q^{2}/\Lambda_{1}+
Q^{4}/\Lambda_{2}+...}\cong \frac{\tilde{g}(0)}{1-Q^{2}/\Lambda_{1}}
\end{equation}
The CMM corrections are introduced by using the equality
\begin{equation}
{\cal N}^{\mu}N_{fi} = J(z)\int \frac{d^{3}ld^{3}l'}{4\omega \omega'}
\varphi^{*}_{P_{f}}(\vec{l'},\omega') \varphi_{P_{i}} (\vec{l}, \omega)\cdot
\langle \vec{l'}|J^{\mu}_{el.mg.} (z)e^{i k z}|\vec{l}\rangle
\end{equation}
\begin{displaymath}
{\cal N}^{1} = \int \frac{d^{3}l}{(4\omega \omega^{'})^{3/2}}
\varphi^{*}_{P_{f}}
(\vec{l}-\vec {k},\omega') \varphi_{P_{i}} (\vec{l}, \omega)\frac{i}
{\sqrt{2}} g(q^{2})
[2\omega(1+\frac{M_{f}}{M_{i}})
|\vec{P_{i}}| -2(\omega-\omega')|\vec{l}| cos \theta]
\end{displaymath}
Here
\begin{equation}
\begin{array}{c}
\omega'^{2}=(\vec{l}-\vec{k})^{2} + M^{2}_{f}  \\ \nonumber
cos \theta = \vec{P_{i}}\cdot \vec{l} / |\vec{P_{i}}| |\vec{l}| \\
{q } = (\omega'-\omega, \vec{k}) \\ \nonumber
\end{array}
\end{equation}
In (6.7) one must introduce the form (6.6) for $g(q^{2})$ . Then using
the equality (6.7), where $\cal {N}^{\mu}$ is determined by the integration
over the model wave functions (6.4), one can determine $g(0)\equiv g $.

Models predictions, based on the parameters listed in Table I are
shown in Table III. The decay widths (6.3) are calculated using either
$\tilde{g}(0) , \ (\tilde{\Gamma})$ or $g(0) , \ (\Gamma)$.
No attempts have been made to select model parameters in order to improve
the agreement with the measured value $\Gamma(\Psi\rightarrow \eta_{c}
\gamma)=(1.12 \pm 0.35) 10^{-6}GeV$ [34].
It is interesting that such, unadjusted, results are in a very reasonable
agreement with the unadjusted results of Ref.[33] (Their Table II, columns
2,3), which were obtained in a quite different quark model.

        The main aim here was to calculate the magnitude of CMM corrections.
They turned out to be $4.4$\%  or smaller, decreasing with the increase of the
heavy quark mass. With $b$ quark present CMM corrections are practically
negligible. Indeed, when one of the valence quarks is very heavy CF and CM
almost coincide [6], so that the spurious CMM almost vanishes.

\newpage

\section{Heavy quark symmetry limit and meson\newline
       decay constants}
\vspace{2cm}
\setcounter{equation}{0}
\renewcommand{\theequation}{\arabic{section}.\arabic{equation}}

\indent

The decay constant $f_{m}$, for a meson $m$, can be calculated
in BCCM [22,24]. The Lorentz covariant CMM corrections are introduced through
equality
\begin{displaymath}
\int d^{4}y J(z)\langle 0|\overline{\Psi} (z^{P}) \gamma^{\mu}\gamma_{5} \Psi
(z^{P})|m,P,M,s=0,y \rangle e^{i q x} = (2\pi)^{4} \delta^{(4)} (q - P)
Z^{\mu} (P)=
\end{displaymath}
\begin{equation}
=\int d^{4}y J(z) \int \frac{d^{3}l}{2\omega} \varphi_{P}(\vec{l},\omega)
\langle 0| J^{\mu}_{5} (z)|l,0,m \rangle \cdot e^{i( q -P)y} e^{i
q z}
\end{equation}

In the r.h.s. of (7.1) goes the meson decay constant $f_{m}$ defined
for a momentum eigenstate $|P,0,m \rangle$
\begin{equation}
< 0| J_{\mu 5} (x)|P,0,m \rangle = \frac {i}{\sqrt {2E(2\pi)^{3}}}
P_{\mu} f_{m} e^{-i P x}
\end{equation}
In the HO potential version of BCCM integrations in (7.1) can be carried out
explicitly. One finds
\begin{displaymath}
\sqrt{6} \frac{P^{\mu}}{M} K_{\alpha,\beta} = \frac{(2\pi)|^{3/2}}{2M
\sqrt{2E}} \varphi_{P}(\vec{P},E) P^{\mu} f_{m}
\end{displaymath}
\begin{equation}
f_{m} = \sqrt{\frac{12}{M_{m} (2\pi)^{3/2} R^{3}_{\alpha,\beta}
C^{(1)}_{\alpha \beta}}} K_{\alpha \beta=m}
\end{equation}
Here
\begin{displaymath}
K_{\alpha\beta} = 4\pi\int dr r^{2}(U_{\alpha}U_{\beta}-V_{\alpha}V_{\beta})
\end{displaymath}
\begin{equation}
c^{(1)}_{\alpha \beta} = 1-3 (\frac{c\alpha}{R^{2}_{0\alpha}}+\frac{c_
{\beta}}{R^{2}_{0\beta}}) R^{2}_{\alpha\beta}+15 \frac{c_{\alpha}c_{\beta}}
{R^{2}_{0\alpha}R^{2}_{0\beta}} R^{4}_{\alpha\beta}
\end{equation}
\begin{displaymath}
R^{2}_{\alpha\beta} = \frac{2R^{2}_{0\alpha}R^{2}_{0\beta}}{R^{2}_{0\alpha}+
R^{2}_{0\beta}} \ \ \ ; \ \ c_{\alpha} =
\frac{\beta_{\alpha}}{4+6\beta^{2}_{\alpha}}
\end{displaymath}
In the HQL (4.5) one has
\begin{displaymath}
K_{\alpha\beta} \rightarrow  4 \pi \int dr r^{2} U_{\alpha}U_{Q} = K_{\alpha_{H
Q L}}
\end{displaymath}
\begin{equation}
R_{\alpha \beta} \rightarrow R_{\alpha_{HQL}}
\end{equation}
\begin{displaymath}
f_{m} \rightarrow \frac{1}{\sqrt{M_{m}}} F_{\alpha_{HQL}} \\ \nonumber
\end{displaymath}
Here $Q$ is a heavy quark $(c,b)$ while $\alpha$  denotes a light quark
(u,d,s).
With (7.5) a meson decay constant has $M^{-1/2}_{m}$ dependence as required by
the heavy quark symmetry (HQS). One obtains for example
\begin{equation}
f_{B_{H Q L}} = \sqrt{\frac{M_{D}}{M_{B}}} f_{D_{HQL}} = 0.6
f_{D_{HQL}}
\end{equation}
With full expression (7.1), using parameters listed in Table I, one obtains
\begin{equation}
\begin{array}{c}
f_{D} = 130.6 MeV \ \ \ ; \   f_{B} = 90.9 MeV \\
f_{B}/f_{D} = 0.696 \\
\end{array}
\end{equation}

The ratio $f_{B}/f_{D}$ (7.7) is in a very good agreement with the result
$f_{B}/f_{D}\cong 0.69$ obtained by the $1/m_{Q}$ expansion of the heavy-light
currents [14,35]. However it is about 30 \% smaller than the results based
on QCD sum rules, lattice calculations and semilocal parton-hadron
duality[36].

The BCCM based calculation gives
\begin{equation}
\begin{array}{c}
f_{D_{s}}=149,2 MeV \\
f_{D_{s}}/f_{D}=1.14 \\
\end{array}
\end{equation}
The ratio $f_{D_{s}}/f_{D}$ is in reasonable agreement with previous results
obtained from lattice QCD or potential models [36]. QCD sum rule analyses
gave $f_{D_{S}}/f_{D}\cong 1.19$ [26] and $f_{D_{S}}/f_{D}\cong 1.1$ [37].
However absolute values (7.7,7.8) for heavy meson decay constants seem to be
smaller than the QCD sum rule or lattice QCD based estimates
[14,26,37-39].

The BCCM with CMM corrections predicts
\begin{displaymath}
f_{K^{+}} = 171 MeV
\end{displaymath}
which is in good agreement with the experimental value $f_{K^{+}}=(160.6 \pm
1.3)MeV$ [34]. The pion decay constant $f_{\pi}=271 MeV$ is to large
$(f_{\pi_{exp}}=(131.73 \pm 0.15)MeV\ [34])$,  as it is usual in valence quark
models.

\newpage

\section{Meson decay form factors and HQS}
\vspace{2cm}
\setcounter{equation}{0}
\renewcommand{\theequation}{\arabic{section}.\arabic{equation}}

\indent

The calculation of meson decay form factors has already been
described [3,5] so only some examples need to be shown here. Matrix elements
for $B\rightarrow D(D^{*})$ transitions are:
\begin{equation}
\begin{array}{c}
\langle P_{f},s=0, D|\overline{c}\gamma^{\mu}b|B,s=0, P_{i}\rangle = \\
= \frac{2\pi\delta^{(4)}(P_{f}+Q-P_{i})}{2\sqrt{E_{i}E_{f}}}
[f_{+}(Q^{2})(P_{i}+P_{f})^{\mu}+f_{-}(Q^{2})(P_{i}-P_{f})^{\mu}] \\
\end{array}
\end{equation}
\begin{equation}
\begin{array}{c}
\langle P_{f},\epsilon, D^{*}|\overline{c}\gamma^{\mu}b|B,s=0, P_{i}\rangle= \\
=\frac{2\pi\delta^{(4)}(P_{f}+Q-P_{i})}{2\sqrt{E_{i}E_{f}}} i g(Q^{2})
\epsilon^{\mu\nu\rho\sigma}\epsilon^{*}_{\nu}(P_{i}+P_{f})_{\rho}(P_{i}-P_{f})
_{\sigma} \\
\end{array}
\end{equation}
\begin{equation}
\begin{array}{c}
\langle P_{f},\epsilon,
D^{*}|\overline{c}\gamma^{\mu}\gamma_{5}b|B,s=0,P_{i}\rangle= \\
\nonumber
=\frac{2\pi\delta^{(4)}(P_{f}+Q-P_{i})}{2\sqrt{E_{i}E_{f}}} [f(Q^{2})
\epsilon^{* \mu}+a_{+}(Q^{2})(\epsilon^{*}\cdot P_{i})(P_{i}+P_{f})^{\mu}+
\\
 + a_{-}(Q^{2})(\epsilon^{*}\cdot P_{i})(P_{i}-P_{f})^{\mu}] \\
\end{array}
\end{equation}
Corresponding BCCM expressions in the generalised Breit frame (6.5) are
\begin{displaymath}
f_{+} = \frac{1}{\sqrt{4M_{i}M_{f}}}[(M_{i}+M_{f})\frac{M_{f}}{E_{f}}
I^{0}_{c b}-(M_{i}-M_{f})\frac{M_{f}}{|\vec {P_{f}}|} I^{3}_{cb}]
Z_{\overline{d}}  \\
\end{displaymath}
\begin{displaymath}
f_{-} = f_{+}[(M_{i}+M_{f})\leftrightarrow (-)(M_{i}-M_{f})] \\
\end{displaymath}
\begin{displaymath}
g = \frac{M_{f}\sqrt{M_{f}M_{i}}}{2 M_{i}E_{f}|\vec {P_{f}}|} V^{1}_{cb}
(\lambda  =+1) Z_{\overline {d}} \\
\end{displaymath}
\begin{equation}
f=\sqrt{4 M_{i}M_{f}}A^{1}_{cb}(\lambda=+1)Z_{\overline d} \\
\end{equation}
\begin{displaymath}
a_{+}=\frac {1}{4M_{i}M_{f}} \sqrt{\frac{M_{f}}{M_{i}}} \{ (M_{i}-M_{f})
(\frac{M_{f}}{|\vec{P_{f}}|})^{2} [\frac{M_{f}}{E_{f}} A^{3}_{c b}
(\lambda=0)-A^{1}_{c b} (\lambda=+1)] + \\
\end{displaymath}
\begin{displaymath}
 + (M_{i}+M_{f}) (\frac {M_{f}}{E_{f}})^{2} [\frac{M_{f}}{|\vec {P_{f}}|}
A^{0}_{cb}(\lambda=0)-A^{1}_{cb}(\lambda=+1)] \} Z_{\overline{d}} \\
\end{displaymath}
\begin{displaymath}
a_{-}=a_{+}[(M_{i}+M_{f})\leftrightarrow (-)(M_{i}-M_{f})] \\
\end{displaymath}
Here

\newpage

\begin{displaymath}
I^{0}_{cb}=4\pi \frac{M_{f}}{E_{f}}\int dr r^{2} j_{0}(\tilde{\rho})
[U_{c}U_{b}+V_{c}V_{b}] \\
\end{displaymath}
\begin{displaymath}
I^{3}_{cb} = 4 \pi \frac{M_{f}}{E_{f}}\int dr r^{2} j_{1} (\tilde{\rho})
[U_{c}V_{b}-V_{c}U_{b}] \\
\end{displaymath}
\begin{displaymath}
V^{1}_{c b}(\lambda=+1) = 4\pi \int dr r^{2}\{ \frac{|\vec{P_{f}}|}{E_{f}}
[{j}_{0}(\tilde{\rho})U_{c}U_{b} - (\frac{1}{3} j_{0}(\tilde{\rho})- \\
\end{displaymath}
\begin{displaymath}
- \frac{2}{3}j_{2}(\tilde{\rho}))V_{c}V_{b}]+j_{1}(\tilde{\rho})[U_{c}V_{b}+
V_{c}U_{b}]\} \\
\end{displaymath}
\begin{equation}
A^{1}_{c b}(\lambda=+1)=4\pi\int dr r^{2}[j_{0}(\tilde{\rho})U_{c}U_{b}-(\frac
{1}{3}j_{0}(\tilde{\rho})-\frac{2}{3} j_{2} (\tilde{\rho}))V_{c}V_{b}+ \\
\end{equation}
\begin{displaymath}
 + \frac{|\vec{P_{f}}|}{E_{f}} j_{1}(\tilde {\rho}) (U_{c}V_{b} + V_{c}U_{b})]
\\
\end{displaymath}
\begin{displaymath}
A^{0}_{cb}(\lambda=0) = - I^{3}_{cb} \\
\end{displaymath}
\begin{displaymath}
A^{3}_{cb}(\lambda=0) = 4\pi\frac{M_{f}}{E_{f}} \int dr r^{2}[j_{0}(\tilde
{\rho})(U_{c}U_{b}-V_{c}V_{b}) + \\
\end{displaymath}
\begin{displaymath}
2(\frac {1}{3}j_{0}(\tilde{\rho})- \frac{2}{3}j_{2}(\tilde{\rho}))V_{c}V_{b}
+\frac {|\vec{P_{f}}|}{M_{f}} j_{1}(\tilde{\rho})U_{c}V_{b}] \\
\end{displaymath}

CMM corrections have been neglected. For heavy-light quark combination
they are always smaller than $5$\% (See Table III). In (8.5) one has introduced
the spherical Bessel functions $j_{l}(\tilde{\rho})$ where:

\begin{equation}
\tilde{\rho}=\frac{M_{f}}{E_{f}} \cdot B_{c b} \cdot |\vec {r}| \\
\end{equation}
\begin{displaymath}
B_{c b}=[(M_{f}+M_{i})-(\epsilon_{c}+\epsilon_{b})]\frac{|\vec{P_{f}}|}{M_{f}}
\end{displaymath}
The symbol $\lambda$ labels the polarization of the vector meson $D^{*}$. The
expressions (8.4) contain also the overlap (free-line) (2.9) of the light
spectator quark.
\begin{equation}
Z_{\overline{d}}=\frac{M_{f}}{E_{f}} 4\pi \int dr r^{2} j_{0}
(\rho)[U^{2}_{d}+V^{2}_{d}] \\
\end{equation}
\begin{displaymath}
\rho =  2 \epsilon_{d} \frac{|\vec{P_{f}}|}{E_{f}}
|\vec{r}| \\
\end{displaymath}

Formulae (8.5) are a version of the more general formulae listed in Appendix
$((A1)-(A7))$ of Ref. [5]. Such formulae are valid for any BCCM, which
includes  BBM [5].

In the HQL (4.6):
\begin{displaymath}
V_{\alpha}=0  \ \ \ \ ; \ \  \alpha = b,c  \\
\end{displaymath}
\begin{displaymath}
U_{b}=U_{c}=U_{HQL}  \\
\end{displaymath}
\begin{displaymath}
B_{cb}\rightarrow 0 \ \ \ \ ; \ \  \tilde{\rho}\rightarrow 0 \ \ \ \ ; \ \
j_{0}(0)=1 \\
\end{displaymath}
\begin{equation}
I^{0}_{HQL} = \frac{M_{f}}{E_{f}} 4\pi \int dr r^{2} U^{2} = \frac
{M_{f}}{E_{f}} K_{HQL} \\
\end{equation}
\begin{displaymath}
I^{3}_{HQL} = 0 \\
\end{displaymath}
\begin{displaymath}
V^{1}_{HQL} = \frac{|\vec{P_{f}}|}{M_{f}} I^{0}_{HQL} \ \ \ \ ; \ \
A^{1}_{HQL} = \frac{E_{f}}{M_{f}}I^{0}_{HQL} \\
\end{displaymath}
\begin{displaymath}
A^{0}_{HQL}= 0 \ \ \ \ \ ; \ \ \  A^{3}_{HQL} = I^{0}_{HQL}
\end{displaymath}

With
\begin{equation}
\begin{array}{c}
|\vec{P_{f}}|^{2}/M^{2}_{f}= \frac{(M_{i}-M_{f})^{2}-Q^{2}}{4 M_{i} M_{f}}
\\ \nonumber
4 E_{i}E_{f} = (M_{i}+M_{f})^{2}[1-\frac {Q^{2}}{(M_{i}+M_{f})^{2}}] \\
\end{array}
\end{equation}

one finds
\begin{displaymath}
f_{+}=\frac{2\sqrt{M_{i}M_{f}}}{(M_{i}+M_{f})}[1-\frac
{Q^{2}}{(M_{i}+M_{f})^{2}}]^{-1} (Z_{\bar{d}}K_{HQL}) \\
\end{displaymath}
\begin{equation}
g=\frac{1}{(M_{i}+M_{f})} f_{+} \\
\end{equation}
\begin{displaymath}
a_{+} = - g \\
\end{displaymath}
\begin{displaymath}
f=2\sqrt{M_{i}M_{f}} (Z_{\overline{d}} K_{HQL}) \\
\end{displaymath}
Very elegant relations among form factors can be found by using the formfactors
from Ref.[18]., i.e.:
\begin{displaymath}
F_{1}=f_{+}  \ \ \ ;\  V=(M_{i}+M_{f})g \\
\end{displaymath}
\begin{equation}
A_{2}=-(M_{i}+M_{f}) a_{+} \\
\end{equation}
\begin{displaymath}
A_{1}=\frac{1}{(M_{i}+M_{f})} f \\
\end{displaymath}
As in HQL $M_{D^{*}}\equiv M_{D}=M_{f}$ one immediately obtains the well-known
[14] HQS relations
\begin{equation}
F_{1}(Q^{2})=V(Q^{2})=A_{2}(Q^{2})= [1-\frac{Q^{2}}{(M_{i}+M_{f})^{2}}]^{-1}
A_{1}
= \frac{2\sqrt{M_{i}M_{f}}}{(M_{i}+M_{f})} \frac{1}{[1-\frac
{Q^{2}}{(M_{i}+M_{f})^{2}}]} Z_{\overline{d}}K_{HQL}
\end{equation}
{}From (8.12) one easily extracts the Isgur-Wise function [7,14]
which is actually determined by the overlap $Z_{\overline{d}}$ (8.7). First one
must
realize that $K_{HQL}$ is actually the HQL of the normalization integral
\begin{equation}
N=\int dr r^{2}(U^{2}+V^{2}) \ \ \ ;\  N_{HQL}=K_{HQL}=1 \\
\end{equation}
Then, with (8.12), (8.13) and the definition [14]
\begin{displaymath}
\xi(v\cdot v')=\lim_{m_{Q}\to \infty} R F_{1}(Q^{2})
\end{displaymath}
\begin{equation}
v\cdot v' = \frac{M^{2}_{i}+M^{2}_{f}-Q^{2}}{2M_{i}M_{f}}
\end{equation}
\begin{displaymath}
R = \frac{2 \sqrt{M_{i}M_{f}}}{M_{i}+M_{f}}
\end{displaymath}
one obtains
\begin{equation}
\xi(v\cdot v')=\frac{4M_{i}M_{f}}{(M_{i}+M_{f})^{2}} \frac{1}{[1-\frac
{Q^{2}}{(M_{i}+M_{f})^{2}}]} Z_{\bar{d}}(Q^{2})
\end{equation}
It should be noted that both (8.12) and (8.15) include explicitly the
kinematic factor $[1-Q^{2}/(M_{i}+M_{f})^{2}]$. Furthermore, at the maximum
momentum transfer $Q^{2}_{max}=(M_{i}-M_{f})^{2}$ one finds [14,40]:
\begin{equation}
\begin{array}{c}
Z_{\overline {d}}|_{\vec {P_{i}} = \vec {P_{f}} = 0} = 1 \\ \nonumber
F_{1}=V=A_{2}=A^{-1}_{1}=R^{-1} \\
\end{array}
\end{equation}

        Thus in HQL the BCCM based relations coincide exactly with QCD
based ones, what is only approximately true for other models [18,29,4].

It might be interesting to compare BCCM prediction for the
$Q^{2}$-dependence of form factors, including the HQL limit, with other
approaches. The results obtained for HO model are shown in Fig. 2 using the
same scale as in corresponding Fig.'s 1.3 and 5.8 in Ref.[14].

        All results presented here can be obtained also in BCCM based on the
MIT-bag model [15]. Fig.3 shows that both versions of BCCM produce
quite simmilar results. Models prediction stay close to the HQS limit, which
is, up to factor $R^{-1}$, given by Isgur-Wise function $\xi(v\cdot
v')$. In BCCM one always obtains $V(Q^{2})> A_{2}(Q^{2})\cong
[1-\frac{Q^{2}}{(M_{B}+M_{D})^{2}}]^{-1} A_{1}> F_{1}$. This ordering
differs from other quark models[14]. It does agree with QCD sum rule
results (Fig. 5.8 in Ref. [14]). However quantitative agreement is not so
good. The absolute values of QCD-sum rule formfactors are usualy
larger than the corresponding BCCM values. The gaps separating  $V,
\ [1-\frac{Q^{2}}{(M_{B}+M_{D})^{2}}]^{-1} A_{1},
\ A_{2}$ and $F_{1}$ curves are also larger. BCCM, as used here, does not
take into account the short distance corrections which are
responsible [14] for 50\% of the enhancement of V relative to $F_{1}$ and
$A_{1}$.

        All  BCCM based conclussions seem to be independent of the form of
central confinement [3,5,15-17]. However the precise form of the
$Q^{2}$-dependence might be influenced by the model detailes. Thus the
selection
of the particular version of BCCM could be some kind of fine tunning.

\newpage

\section{Main characteristics}
\vspace{2cm}
\setcounter{equation}{0}
\renewcommand{\theequation}{\arabic{section}.\arabic{equation}}

\indent

        The main aim of this paper was to demonstrate how one can construct
a whole class of quark models which are heavy quark (b,c) symmetric. Such
models are also Lorentz covariant, as it has been shown in Ref's [3] and [5].
The kinematic factor (8.12), (8.15) which appears in HQS relations is a
typical consequence of the Lorentz covariance.

        The class of HQS models contains models [15-17] in
which each quark is independently centrally confined. As it is well
known [25] such models experience spurious CMM effects. It is demonstrated
here that one can introduce CMM corrections in the manifestly covariant way.
They notably improove $\mu_{p}, g_{A}$, hadron masses and other quantities
which involve "light" (u,d,s) quarks. For "heavy-light" combinations CMM
corrections diminish with the increase of the heavy quark mass (See Tables
II,III). In the derivation of HQS relations (8.12) they could have been
neglected. However their presence, as in (7.6), does not spoil HQS
character of the model.

        A BCCM is based on a static quark model (Examples in Ref.s 15-17)
with specified boosts (2.3), hyperplane projection (2.12) and overlaps
(2.9). After BCCM is formulated all calculations depend only on the
parameters of the underlaying static model. Basing BCCM  on the MIT-bag
model one uses the usual bag-model parameters [15]. With a harmonic
oscillator potential as a starting point one employs only parameters listed
in Table I. All form factors (7.3), (8.4), HQS relations (7.6), (8.12),
Isgur-Wise function (8.15) etc are obtained by a straigthforward
calculation, without any additional "add hoc" assumptions. The results in
Table II, excellent $g_{A}$ value and reasonable $\mu_{p}$ (5.9), are not due
to any "fine tuning". Playing with parameters one could "improve" some of
those outcomes, which would be pointless, as it does not lead to any new
physical insights. Of more fundamental importance could be the selection of
the type of central confinement. It, obviously, pays to select the
confinement which best mimicks the real physics. Some idea about the
confinement dependence can be obtained by comparison of Fig.'s 2 and 3.

        As it is usual with central valence quark models [15-17], BCCM
also fails in the description of pion, by not beeing able to account
for its Goldstone-boson character.

        BCCM's can be used to calculate corrections ((7.7), Fig.'s 2 and 3)
to the extreme HQS results. However only those corrections which depend on
the valence quark dynamics are included. Short distance QCD
corrections [14] were not incorporated in BCCM. As the model is formulated
in the quantum field formalism (2.5), which can be related to Furry
bound state picture [27,42], some estimates of QCD effects might become
feasible.

        An important characteristic of the class of BCCM's is that those
models describe mesons and baryons within the same formalism. Here BCCM's
were mostly applied to mesons, but calculations of the electromagnetic [3-5]
(5.4) and of the semileptonic [3] baryon form factors are equally feasible.

\newpage
\begin{Large}
\appendix{\bf Appendix}
\end{Large}
\vspace{2cm}
\setcounter{equation}{0}
\renewcommand{\theequation}{A\arabic{equation}}

\indent

The explicit expressions for the components $\phi$ of the momentuum
eigenstates can be found by using (3.8), (4.3) and Table I.
For proton (nucleon) one finds:
\begin{equation}
\frac{M}{\omega_{l}}|\phi_{P}(\vec{l},\omega_{l})|^{2} = \frac{\tilde{\omega_
{l}}}{M}(\frac{R_{0}}{\sqrt{3\pi}})^{3} e^{-\frac{\tilde{\vec{l^{2}}}R^{2}
_{0}}{3}}
 \{C_{0}+C_{1}(\tilde{\vec {l}})^{2}+C_{2}(\tilde{\vec{l}})^{4}+C_{3}
(\tilde{\vec {l}})^{6}\} \\
\end{equation}
Here
\begin{displaymath}
\tilde{\omega}=\frac {P_{\mu}l^{\mu}}{M} \ \ \ \  ;\ \ \tilde{\vec {l}}=\vec
{l} +
\frac {\vec {P}(\vec {P}\vec {l})} {M(E+M)} - \frac {\vec {P}}{M}
\omega_{l}\\
\end{displaymath}
\begin{displaymath}
\tilde{\omega}^{2}-\tilde{\vec {l^{2}}}=M^{2}\\
\end{displaymath}
\begin{displaymath}
C_{0}=1-6c+20c^{2}-\frac {280}{9} c^{3}\\
\end{displaymath}
\begin{equation}
C_{1}=\frac {4}{3}[c-\frac{20}{3}c^{2}+\frac {140}{9} c^{3}]R^{2}_{0}\\
\end{equation}
\begin{displaymath}
C_{2}=\frac {16}{27}[c^{2}-\frac {14}{3} c^{3}]R^{4}_{0}\\
\end{displaymath}
\begin{displaymath}
C_{3}=\frac {64}{729}c^{3}R^{6}_{0}\\
\end{displaymath}
\begin{displaymath}
c=\frac {\beta^{2}}{4+6\beta^{2}} \\
\end{displaymath}

        The normalization of the proton component (A1) is
\begin{equation}
\int d^{3}l \frac{M^{2}}{\omega^{2}_{l}}|\phi_{P}(\vec{l},
\omega_{l})|^{2}=1
\end{equation}

The meson component is
\begin{equation}
\frac{|\varphi_{P}(\vec{l},\omega_{l})|^{2}}{2\omega_{l}} =
\frac{2(l\cdot P)}{(2\pi)^{3/2}M} R^{3}_{ab}e^{ -
\tilde{\vec{l}^{2}}R^{2}_{ab}/2}
\cdot [C^{(1)}_{ab} + C^{(2)}_{ab}\tilde{\vec{l^{2}}} +
C^{(3)}_{ab}\tilde{\vec{l^{4}}}]
\end{equation}

Here with flavors a,b:
\begin{displaymath}
\tilde{\omega}=\frac {P_{\mu}l^{\mu}}{M} \ \ \ \  ;
\ \ \tilde{\vec {l}}=\vec {l}+\frac {\vec {P}\cdot \vec {l}}{M(E+M)}-\frac
{\vec
{P}}{M}\omega_{l}\\
\end{displaymath}
\begin{displaymath}
C^{(1)}_{ab} = 1-3(\frac {c_{a}}{R^{2}_{0a}}+\frac {c_{b}}{R^{2}_{0b}})
R^{2}_{ab}+15\frac {c_{a}c_{b}}{R^{2}_{0a}R^{2}_{0b}}R^{4}_{ab}\\
\end{displaymath}
\begin{equation}
C^{(2)}_{ab}=(\frac {c_{a}}{R^{2}_{0a}} + \frac {c_{b}}{R^{2}_{0b}})R^{4}_{ab}
-10 \frac{c_{a}c_{b}}{R^{2}_{0a}R^{2}_{0b}}R^{6}_{ab}\\
\end{equation}
\begin{displaymath}
C^{(3)}_{ab}=\frac {c_{a}c_{b}}{R^{2}_{0a}R^{2}_{0b}}R^{8}_{ab} \\
\end{displaymath}
\begin{displaymath}
R^{2}_{ab}= \frac {2R^{2}_{0a}R^{2}_{0b}}{R^{2}_{0a}+R^{2}_{0b}} \\
\end{displaymath}
\begin{displaymath}
c_{a}=\frac {\beta^{2}_{a}}{4+6\beta^{2}_{a}} \\
\end{displaymath}

The normalization is
\begin{equation}
\int d^{3}l \frac {1}{4\omega^{2}_{l}}|\varphi_{P}(\vec
{l},\omega_{l})|^{2}=1
\end{equation}

\newpage

{\bf Table I.} HO model parameters

\vspace{1cm}

\begin{tabular}{|c|c|c|c|c|}
\hline
Flavor & m(GeV) & E(GeV) & $\beta$ & $R_{o}(GeV^{-1})$ \\ \hline
u,d & 0.315 & 0.426 & 0.455 & 2.96 \\
s & 0.525 & 0.557 & 0.343 & 2.70 \\
c & 1.850 & 1.710 & 0.140 & 2.00 \\
b & 5.450 & 5.221 & 0.062 & 1.52 \\ \hline
\end{tabular}

\newpage

{\bf Table II.} Hadron masses in HO model

\vspace{1cm}

\begin{tabular}{|c|c|c|c|c|c|}
\hline
Hadron & $\tilde{M}^{*}$ & $M^{*}$ & $M_{Exp}^{*}$ & $M/\tilde{M}$ &
 $|\frac{M-M_{exp}}{M_{exp}}|$\% \\ \hline
p & 1.191 & 0.928 & 0.938 & 0.78 & 1.1\\ \hline
$\Delta$ & 1.365 & 1.138 & 1.236 & 0.83 & 8.0\\ \hline
$\pi$ & 0.679 & 0.329 & 0.139 & 0.48 & --  \\ \hline
$\rho$ & 0.910 & 0.677 & 0.770 & 0.74 & 12.1\\ \hline
K & 0.817 & 0.528 & 0.498 & 0.65 & 6.0\\ \hline
$K^{*}$ & 1.019 & 0.798 & 0.892 & 0.78 & 10.5\\ \hline
$\eta_{c}$ & 3.381 & 3.267 & 2.979 & 0.97 & 9.7\\ \hline
$\Psi$ & 3.433 & 3.322 & 3.097 & 0.97 & 7.3\\ \hline
$D^{+}$ & 1.906 & 1.752 & 1.869 & 0.92 & 6.3\\ \hline
$D^{+*}$ & 2.005 & 1.858 & 2.010 & 0.93 & 7.6\\ \hline
$D_{s}$ & 2.138 & 1.994 & 1.969 & 0.93 & 1.2\\ \hline
$D_{s}^{*}$ & 2.229 & 2.091 & 2.110 & 0.94 & 0.9\\ \hline
$B^{+}$ & 5.207 & 5.125 & 5.279 & 0.98 & 2.9\\ \hline
$B^{+*}$ & 5.249 & 5.168 & 5.325 & 0.98 & 2.9\\ \hline
$B_{s}$ & 5.580 & 5.501 & 5.384 & 0.99 & 2.1\\ \hline
$B^{*}_{s}$ & 5.620 & 5.541 & 5.431 & 0.99 & 2.0\\ \hline
\end{tabular}

$^{*}$ All masses are in GeV.

\newpage

{\bf Table III.} The M1 transition decay widths without $(\tilde{\Gamma})$ and
with $(\Gamma)$ CMM corrections

\vspace{1cm}

\newcommand{\sm}[1]{\footnotesize #1}

\begin{tabular}{|c|c|c|c|c|c|}
\hline
MODE & $\sm{\tilde{g}(0) (GeV^{-1})}$ & $\sm{\tilde{\Gamma_{0}}
(10^{-6}GeV)}$ & $\sm{g(0) (GeV^{-1})}$ & $\sm{\Gamma (10^{-6}
GeV)}$ &
$\sm{\frac{g(0)-\tilde{g}(0)}{\tilde{g}(0)}}$\% \\ \hline
$\Psi\rightarrow \eta_{c}\gamma$ & 0.367 & 2.041 & 0.377 & 2.148 & 2.6\\
\hline
$D^{+*}\rightarrow D^{+}\gamma$ & -0.127 & 0.393 & -0.132 & 0.429 & 4.4\\
\hline
$D^{0*}\rightarrow D^{0}\gamma$ & 0.811 & 16.656 & 0.847 & 18.164 & 4.4\\
\hline
$D^{+*}_{s}\rightarrow D^{+}_{s}\gamma$ & -0.062 & 0.094 & -0.064 & 0.103 & 4.2
\\ \hline
$B^{+*}\rightarrow B^{+}\gamma$ & 0.639 & 0.382 & 0.645 & 0.389 & 1.0\\
\hline
$B^{0*}\rightarrow B^{0}\gamma$ & -0.367 & 0.126 & -0.371 & 0.128 & 1.0\\
\hline
$B^{0*}_{s}\rightarrow B^{0}_{s}\gamma$ & 0.290 & 0.084 & 0.293 & 0.086 & 1.0
\\ \hline
\end{tabular}

\newpage

{\bf Figure captions}

\vspace{1cm}

{\bf Fig.1} Vertex for the semileptonic  $B\rightarrow D$ transitions. Quark
lines
$(b,c, \bar {d})$, momenta $P_{i,f}$ and the overlap integral $Z$ are
indicated

{\bf Fig.2} Predictions for the weak decay formfactors in HO based BCCM.
Dot-dashed
line corresponds to $V$, full line to $A_{2}$ and
$[1-Q^{2}/(M_{B}+M_{D})^{2}]^{-1} A_{1}$ and the dashed line to $F_{1}$.
HQS limit coincides with the full line.

{\bf Fig.3} Predictions for the weak decay form factors in MIT-bag based BCCM.
Line identification is the same as in Fig. 2.
\end{document}